\documentclass[times,11pt]{article}
\newcommand{\etal}{{\it et al.\ }}

\newcommand{\beq}{\begin{equation}
  \renewcommand{\int}{\intop\limits}
  \renewcommand{\oint}{\ointop\limits}}
\newcommand{\eeq}{\end{equation}}
\newcommand{\beqarr}{\par\begin{minipage}{11cm} \begin{eqnarray*}}
\newcommand{\eeqarr}{\end{eqnarray*} \end{minipage} \hfill 
   \stepcounter{equation}{\rm (\theequation)}\vspace{3mm}\linebreak}
\newcommand{\bdm}{\begin{displaymath}
  \renewcommand{\int}{\intop\limits}
  \renewcommand{\oint}{\ointop\limits}}
\newcommand{\edm}{\end{displaymath}}

\newcommand{\up}[1]{\ifmmode^{\rm #1}\else$^{\rm #1}$\fi}

\newcommand{\arcd}{\ifmmode^{\circ}\else$^{\circ}$\fi}
\newcommand{\arcm}{\ifmmode{'}\else$'$\fi}
\newcommand{\arcs}{\ifmmode{''}\else$''$\fi}

\newcounter{pagefrom}
\newcounter{pageto}
\newcounter{volume}
\newcounter{year}

\newenvironment{Titlepage}{
\vspace*{2cm}
  \begin{center}
}{
  \end{center}\par\vspace{3mm}
}

\newcommand{\Title}[1]{{\large\bf\boldmath #1 \\[3mm] {\footnotesize by} 
\\[3mm]}}

\newcommand{\Author}[2]{{\large\spaceskip 2pt plus 1pt minus 1pt #1}\\[3mm]
   {\small #2}\\[6mm]}

\newcommand{\Received}[1]{}

\newcommand{\Abstract}[2]{{\footnotesize\begin{center}ABSTRACT\end{center}
\vspace{1mm}\par#1\par
\noindent
{\bf Key words:~~}{\it #2}}}

\newcommand{\FigCap}[1]{\footnotesize\par\noindent Fig.\  % 
  \refstepcounter{figure}\thefigure. #1\par}

\newcommand{\TabCap}[2]{\begin{center}\parbox[t]{#1}{\begin{center}
  \small {\spaceskip 2pt plus 1pt minus 1pt T a b l e}
  \refstepcounter{table}\thetable \\[2mm]
  \footnotesize #2 \end{center}}\end{center}}

\newcommand{\TableFont}{\footnotesize}

\newcommand{\MakeTable}[4]{\begin{table}[htb]\TabCap{#2}{#3}
  \begin{center} \TableFont \begin{tabular}{#1} #4 
  \end{tabular}\end{center}\end{table}}

\newcommand{\MakeTableSep}[4]{\begin{table}[p]\TabCap{#2}{#3}
  \begin{center} \TableFont \begin{tabular}{#1} #4 
  \end{tabular}\end{center}\end{table}}

{
\footnotesize \frenchspacing

\renewcommand{\and}{{\rm and }}
\section{{\rm REFERENCES}}
\sloppy \hyphenpenalty10000
\begin{list}{}{\leftmargin1cm\listparindent-1cm
\itemindent\listparindent\parsep0pt\itemsep0pt}}%
{\end{list}\vspace{2mm}}

\def\TYLDA{~}
\newlength{\DW}
\newcommand{\refitem}[5]{\item[]{#1} #2%
\def\REFARG{#3}\ifx\REFARG\TYLDA\else, {\it#3}\fi
\def\REFARG{#4}\ifx\REFARG\TYLDA\else, {\bf#4}\fi
\def\REFARG{#5}\ifx\REFARG\TYLDA\else, {#5}\fi.}

\usepackage{supertabular,lscape,epsfig}
\usepackage{amssymb}
\usepackage{amsmath}
\DeclareSymbolFont{ppa}{OT1}{ppl}{m}{it}
\DeclareMathSymbol{\vv}{\mathalpha}{ppa}{'166}
\begin{document}

\newcommand{\TabCapp}[2]{\begin{center}\parbox[t]{#1}{\centerline{
  \small {\spaceskip 2pt plus 1pt minus 1pt T a b l e}
  \refstepcounter{table}\thetable}
  \vskip2mm
  \centerline{\footnotesize #2}}
  \vskip3mm
\end{center}}

\newcommand{\TTabCap}[3]{\begin{center}\parbox[t]{#1}{\centerline{
  \small {\spaceskip 2pt plus 1pt minus 1pt T a b l e}
  \refstepcounter{table}\thetable}
  \vskip2mm
  \centerline{\footnotesize #2}
  \centerline{\footnotesize #3}}
  \vskip1mm
\end{center}}

\newcommand{\MakeTableSepp}[4]{\begin{table}[p]\TabCapp{#2}{#3}
  \begin{center} \TableFont \begin{tabular}{#1} #4
  \end{tabular}\end{center}\end{table}}

\newcommand{\MakeTableee}[4]{\begin{table}[htb]\TabCapp{#2}{#3}
  \begin{center} \TableFont \begin{tabular}{#1} #4
  \end{tabular}\vspace*{-7mm}\end{center}\end{table}}

\newcommand{\MakeTablee}[5]{\begin{table}[htb]\TTabCap{#2}{#3}{#4}
  \begin{center} \TableFont \begin{tabular}{#1} #5
  \end{tabular}\end{center}\end{table}}

\newcommand{\FigurePs}[7]{\begin{figure}[htb]\vspace{#1}
\includegraphics{#4}
\FigCap{#2\label{#3}}
\end{figure}}

\newfont{\bb}{ptmbi8t at 12pt}
\newfont{\bbb}{cmbxti10}
\newfont{\bbbb}{cmbxti10 at 9pt}
\newcommand{\uprule}{\rule{0pt}{2.5ex}}
\newcommand{\douprule}{\rule[-2ex]{0pt}{4.5ex}}
\newcommand{\dorule}{\rule[-2ex]{0pt}{2ex}}
\def\thefootnote{\fnsymbol{footnote}}

\begin{Titlepage}
\Title{The All Sky Automated Survey. The Catalog of Variable Stars.
V. Declinations 0\arcd -- 28\arcd of the Northern Hemisphere}
\Author{G.~~P~o~j~m~a~{\'n}~s~k~i,~~~B.~~P~i~l~e~c~k~i,~~~D.~~S~z~c~z~y~g~i~e~{\l}}{Warsaw University Observatory,
Al~Ujazdowskie~4, 00-478~Warszawa, Poland\\
e-mail:(gp, pilecki, dszczyg)@astrouw.edu.pl}
\end{Titlepage}

\Abstract{This paper contains the fifth part of the
Catalog of Variable Stars created from the $V$-band photometric data
collected  by  $9\arcd \times 9\arcd$ camera of the All Sky Automated Survey.
Preliminary list of variable stars found in the fields located
between declination 0\arcd and +28\arcd of the northern hemisphere 
is presented. 
11,509 stars brighter than $V$=15 were found to be
variable (2,482 eclipsing, 1,397 regularly pulsating, 318 Mira and
7,310 other stars). Automated algorithm taking into account light curve
properties (period, fourier coefficients) and other
available data (2MASS colors, IRAS fluxes) was applied to 
roughly classify objects.  Basic photometric properties are
presented in the tables and thumbnail light
curves are made available for reference.
All the photometric data are  available over the INTERNET at
{\it http://www.astrouw.edu.pl/\~{}gp/asas/asas.html} or {\it http://archive.princeton.edu/\~{}asas}}{Catalogs --Stars: variables: general -- Surveys}

\section{Introduction}
The All Sky Automated Survey (ASAS, Pojma\'nski 1997) is a
photometric CCD sky survey monitoring entire southern 
and part of the northern sky ($\delta<25\arcd$)
since October 2000.
It was triggered by ideas of Paczy\'nski (1997) of using small 
automated instruments for bright star surveys.

The ASAS system (Pojma\'nski 2001), 
located in Las Campanas Observatory (operated
by the Carnegie Institution of Washington), consists of four
instruments equipped with two wide field ($9\arcd \times 9\arcd$,
$B$ and $V$ filters), one very wide ($36\arcd \times 36\arcd$, $R$ filter) and
one narrow field ($2\arcd \times 2\arcd$, $I$ filter) cameras, with
2K$\times$2K CCD attached.
This configuration allows us to obtain photometry of all
sources brighter than limiting magnitude $V\sim~14$ ($I\sim~13$).

Variability analysis was performed using the $V$-band data as soon as
reasonable amount of data had been collected. We have already
presented preliminary catalogs of variable stars in the Southern Hemisphere
(Pojma\'nski 2002, 2003, Pojma\'nski and Maciejewski 2004, 2005)
This paper contains the fifth part of the analyzed data - variable stars
located in the fields located in the Northern Hemisphere ($\delta < +28\arcd$).

\section{Observations and Data Reduction}

Data acquisition and processing in the ASAS system is fully automated, 
starting from taking calibration frames (BIAS, DARK, FLAT), scheduling
observations and telescope control, through image processing,
photometry, astrometry, data transfer and backup, to 
incorporating new measurements into the ASAS Photometric Catalog.

The data reduction pipe-line used to process ASAS data was
described in details by Pojma\'nski (1997). 
We are making simultaneous photometry through five
apertures (2 to 6 pixels in diameter). Each aperture data is processed
separately. Data obtained with the smallest one is used
for the faint ($V > 12$) stars and with the largest one for the
bright ($V < 9$) objects.
Profile fitting and image subtraction were tested as an alternative,
 but did not perform well with highly variable and
undersampled ASAS images.

Astrometric calibration is currently based on the ACT
(Urban \etal 1998) catalog. Typical positional accuracy is
around 0.2 pixels ( $\sim3$~arc sec).

The zero-point offset of photometry is based on the Tycho
(Perryman {\em et al.} 1997) data and in most areas in the sky 
is accurate to about 0.05 mag. However, due to the non-perfect
flat-fielding, missing color information and blending
of the stars, much larger errors can be sometimes observed. Differential
accuracy is much better, reaching 0.01 for bright stars.

\section{Variability Search and Classification}

Variability analysis was the same as the one used to complete previous
parts of this catalog (Pojma\'nski 2000). Only 
top 5 \% of the stars showing large dispersion of measurements 
were analyzed. Light curves were tested for periodicity with
the Analysis Of Variance test (Schwarzenberg-Czerny 1989). Non-periodic 
behavior was detected with dedicated long-term variability test.
All detections were inspected and confirmed visually.

The automated classification algorithm, described in details in 
previous parts of this catalog (Pojma{\'n}ski 2002, 2003, 2004), 
consists of a few basic steps:
First, 2MASS and IRAS counterparts are identified, providing
$J,H,K$ for all stars (with 10\% abiguity due to blending) 
and infrared fluxes for some of them.
Strictly periodic variables are then classified into predefined classes
using periods, amplitudes, Fourier coefficients of the light curves, 
$H-K$ and  $J-H$ colors and infrared fluxes.
Less periodic light curves and other unusual objects
are simply assigned MISC type.

\section{The Catalog}

We have selected 11,509 variable stars in the equatorial zone of the Northern 
Hemisphere.

For each star the following data are provided: ASAS identification $ID$
(coded from the star's $\alpha_{2000}$ and $\delta_{2000}$ in the
form: $hhmmss-ddmm.m$), period $P$ in days (or characteristic time
scale of variation for irregular objects), $T_0$ -  epoch of minimum
(for eclipsing) or maximum (for pulsating) brightness, $V_{max}$ -
brightness at maximum, $\Delta V$ - amplitude of variation, $Type$ - one
of the predefined classes: $DSCT$, $RRC$, $RRAB$, $DCEP_{FU}$, $DCEP_{FO}$,
$CW$, $ACV$, $BCEP$, $M$ and $MISC$. GCVS cross-identification,
$J$, $J-H$ and $H-K$ data taken from 2MASS catalog are also provided.

\tabcolsep 5pt
\MakeTable{|l|r|l|r|}{8cm}{\label{tabvar}
Number of various types of variable stars detected between 
declinations 0\arcd and +28\arcd of the Northern Hemisphere 
by the ASAS-3 $V$ camera.}{
\hline
\multicolumn{1}{|c|}{Type} & \multicolumn{1}{c|}{Count} & \multicolumn{1}{c|}{Type} & \multicolumn{1}{c|}{Count}\\
\hline
$DCEP_{FU}$ & 136   & $DSCT$&  435 \\
$DCEP_{FO}$ &  46   & $EC$  &  1283\\
$CW$        &  172   & $ED$  & 433 \\
$ACV$       &  65   & $ESD$ & 766 \\
$BCEP$      &  15   & $M$   & 318 \\
$RRAB$      & 343   & $MISC$& 7310 \\
$RRC$       & 185   &       &  \\
\hline
}

Search for GCVS (Kholopov, \etal 1985) variables revealed about 2,402 possible
matches within 3 arc minute radius.

Numbers of stars that have been primarily classified as a given type are listed in Table 1.
Table 2. contains a compact version of the
catalog. Only four  columns are
listed  for each star: identification $ID$, $P$, $V_{max}$, and $\Delta V$.
Column $ID$ also contains some
flags - ":" if classification was uncertain, "?" if multiple classes
were assigned (objects were grouped in the table according to the highest rank
assignment), "v" if GCVS data for this object exist.

Appendix contains thumbnail plots of the light curves identified by their 
$ID$'s.
For periodic variables
phase in the range ($-0.1$ - $2.1$) is plotted along the $x$-axis, while for
Mira's and miscellaneous stars -  HJD in the range (2451800-2453500).
Larger ticks on $y$-axis always mark
1 magnitude intervals and vertical span is never smaller than 1 mag.

The full catalog of variable stars observed by the ASAS system,
containing more classification details, as well as complete data for the
light curves, is available over the INTERNET:\\
\centerline{\it http://www.astrouw.edu.pl/\~{}gp/asas/asas.html}
 or
\centerline{\it http://archive.pinceton.edu/\~{}asas}

\section{Conclusions}

This paper, listing 11,509 objects in the Northern Hemisphere with 
declinations smaller then +28\arcd concludes preliminary search for
variable stars in the All Sky Automated Survey photometric data,
increasing the total number of variables detected by ASAS to
almost 50,000.

Although containing much more uniform data than any other
existing database, the ASAS catalog is still incomplete.
Our future efforts will include repeating variability search in the
first quarter of the Southern Hemisphere, incorporating narrow-eclipse 
binaries discovered by the ASAS Alert System (over 2,500 objects)
and search for low-amplitude variables all over the sky.
Test performed just in one of 500 fields revealed that including 
low-amplitude stars will double the size of the ASAS Catalog
of Variable Stars.

\section{Acknowledgments}
This project was made possible by a generous gift from Mr. William
Golden to Dr. Bohdan Paczy{\'n}ski, and funds from Princeton University.  It is a
great pleasure to thank Dr. B. Paczy{\'n}ski for his initiative, interest,
valuable discussions, and the funding of this project.

I am indebted to the OGLE collaboration (Udalski, Kubiak, Szyma{\'n}ski 1997)
for the use of facilities of the
Warsaw telescope at LCO, for their permanent support and maintenance of the
ASAS instrumentation, and to The Observatories of the
Carnegie Institution of Washington for providing
the excellent site for the observations.

This research has made use of the SIMBAD database,
operated at CDS, Strasbourg, France and
of the NASA/ IPAC Infrared Science Archive, which is operated by the Jet Propulsion Laboratory, California Institute of Technology, under contract with the National Aeronautics and Space Administration.

This publication makes use of data products from the Two Micron All Sky Survey, which is a joint project of the University of Massachusetts and the Infrared Processing and Analysis Center/California Institute of Technology, funded by the National Aeronautics and Space Administration and the National Science Foundation.

This work was supported by the KBN 2P03D02024 grant.

\vspace{1.5cm}
\begin{center}
References
\end{center}

\noindent
\begin{itemize}
\leftmargin 0pt
\itemsep -5pt
\parsep -5pt
\refitem{Kholopov, P.N., \etal}{1985}{~}{~}{General Catalog of Variable
Stars, The Fourth Edition, Nauka, Moscow}
\refitem{Paczy\'nski, B.}{1997}{~}{~}{``The Future of Massive
Variability Searches'', in {\it Proceedings of 12th IAP
Colloquium}: ``Variable Stars and the Astrophysical Returns of
Microlensing Searches'', Paris (Ed. R. Ferlet), p.~357}
\refitem{Perryman, M.A.C. \etal}{ 1997}{ Astron. Astroph}{ 323}{
L49}
\refitem{Pojma\'nski, G.}{1997}{Acta Astron.}{47}{467}
\refitem{Pojma\'nski, G.}{2000}{Acta Astron.}{50}{177}
\refitem{Pojma{\'n}ski, G.}{2001}{~}{~}{
``The All Sky Automated Survey (ASAS-3) System - Its Operation and Preliminary 
Data'' ini: Small Telescope Astronomy on Global Scales,
{\it ASP Conference Series} Vol. 246, IAU Colloquium 183. Edited by
Bohdan Paczynski, Wen-Pin Chen, and Claudia Lemme. San Francisco:
Astronomical Society of the Pacific, p. 53}{}{} 
\refitem{Pojma\'nski, G.}{2002}{Acta Astron.}{52}{397}
\refitem{Pojma\'nski, G.}{2003}{Acta Astron.}{53}{341}
\refitem{Pojma\'nski, G., Maciejewski, G.}{2004}{Acta Astron.}{54}{153}
\refitem{Pojma\'nski, G., Maciejewski, G.}{2005}{Acta Astron.}{1}{1}
\refitem{Udalski, A. Kubiak, M., and Szyma\'nski, M.}{1997}{Acta Astr.}{47}{319}
\refitem{Schwarzenberg-Czerny, A.}{1996}{Astrophys. J.}{460}{L107}
\refitem{Urban, S.E., Corbin, T.E., Wycoff, G.L.}{1998}{AJ}{115,1709}{2161}
\end{itemize}

\tabcolsep 3pt
\input{vartab.tex}

\textheight 24cm
\begin{figure}[p]
\vglue-3mm
\centerline{\bf Appendix}
\vskip1mm
\centerline{\bf ASAS Atlas of Variable Stars. Declinations {\bf 0\arcd -- 28\arcd} of the Northern Hemisphere}

\centerline{\small Only several light curves of each type are printed. Full Atlas
 is
available over the {\sc Internet}:}

\centerline{\it http://www.astrouw.edu.pl/\~{}gp/asas/appendixeq.ps.gz}
%% EC ESD ED DSCT BCEP ACV RRC RRAB DCEP-FU DCEP-FO CW
\vskip20mm
\centerline{Stars classified as EC}
\vskip3.5cm
\centerline{Stars classified as ESD}
\vskip3.5cm
\centerline{Stars classified as ED}
\vskip3.5cm
\centerline{Stars classified as DSCT}
\vskip-11.7cm
\centerline{\includegraphics[bb=30 70 400 495, width=13cm]{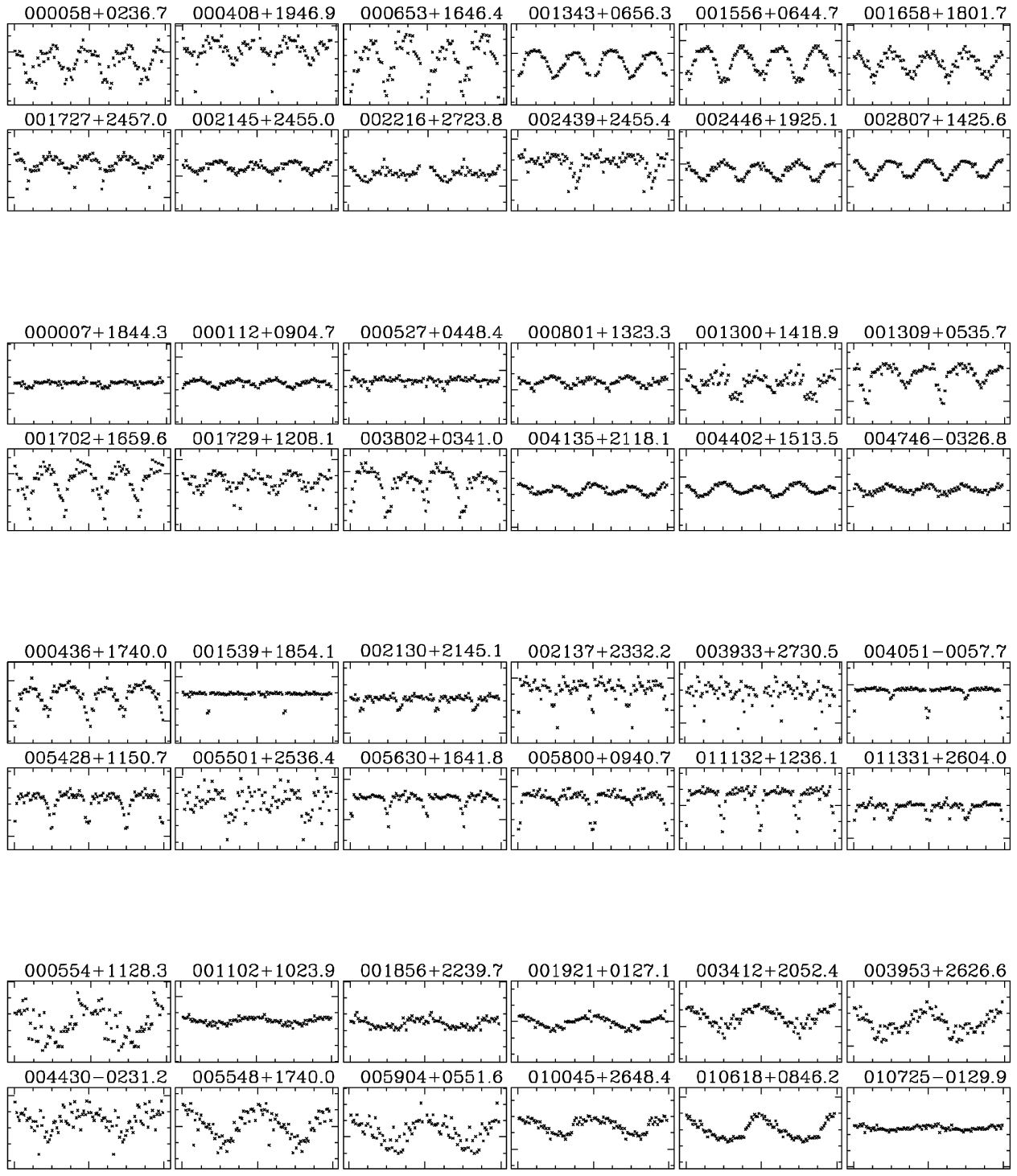}}
\end{figure}
\begin{figure}[p]
\vskip-2mm
\centerline{Stars classified as BCEP}
\vskip3.5cm
\centerline{Stars classified as ACV}
\vskip3.5cm
\centerline{Stars classified as RRC}
\vskip3.5cm
\centerline{Stars classified as RRAB}
\vskip3.5cm
\centerline{Stars classified as DCEP-FU}
\vskip-15.7cm
\centerline{\includegraphics[bb=30 30 405 610, width=13cm]{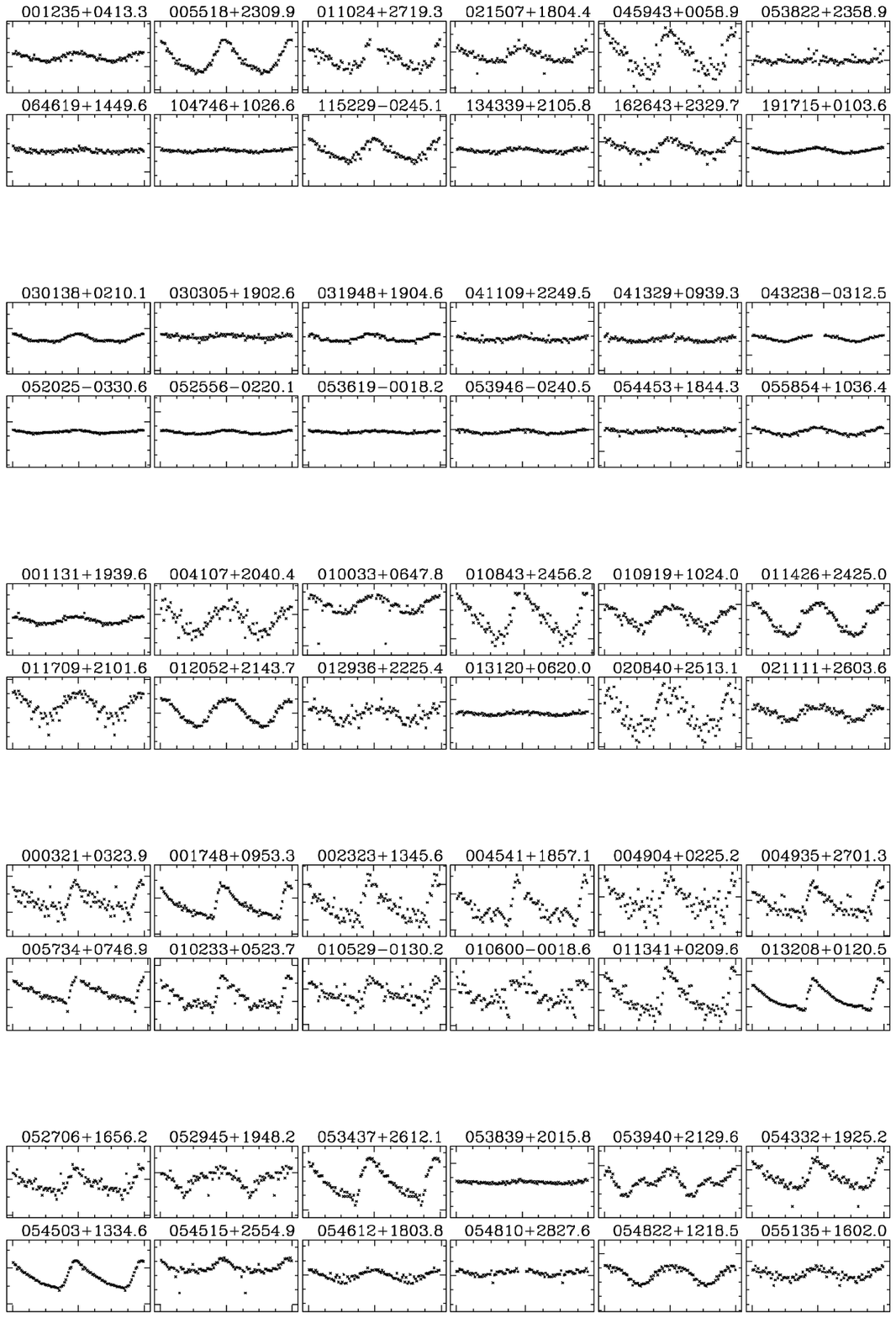}}
\end{figure}

\begin{figure}[p]
\vskip-2mm
\centerline{Stars classified as DCEP-FO}
\vskip3.5cm
\centerline{Stars classified as CW}
\vskip3.5cm
\centerline{Stars classified as M}
\vskip3.5cm
\centerline{Stars classified as MISC}
\vskip3.5cm
\vskip-15.2cm
\centerline{\includegraphics[bb=30 65 405 610, width=13cm]{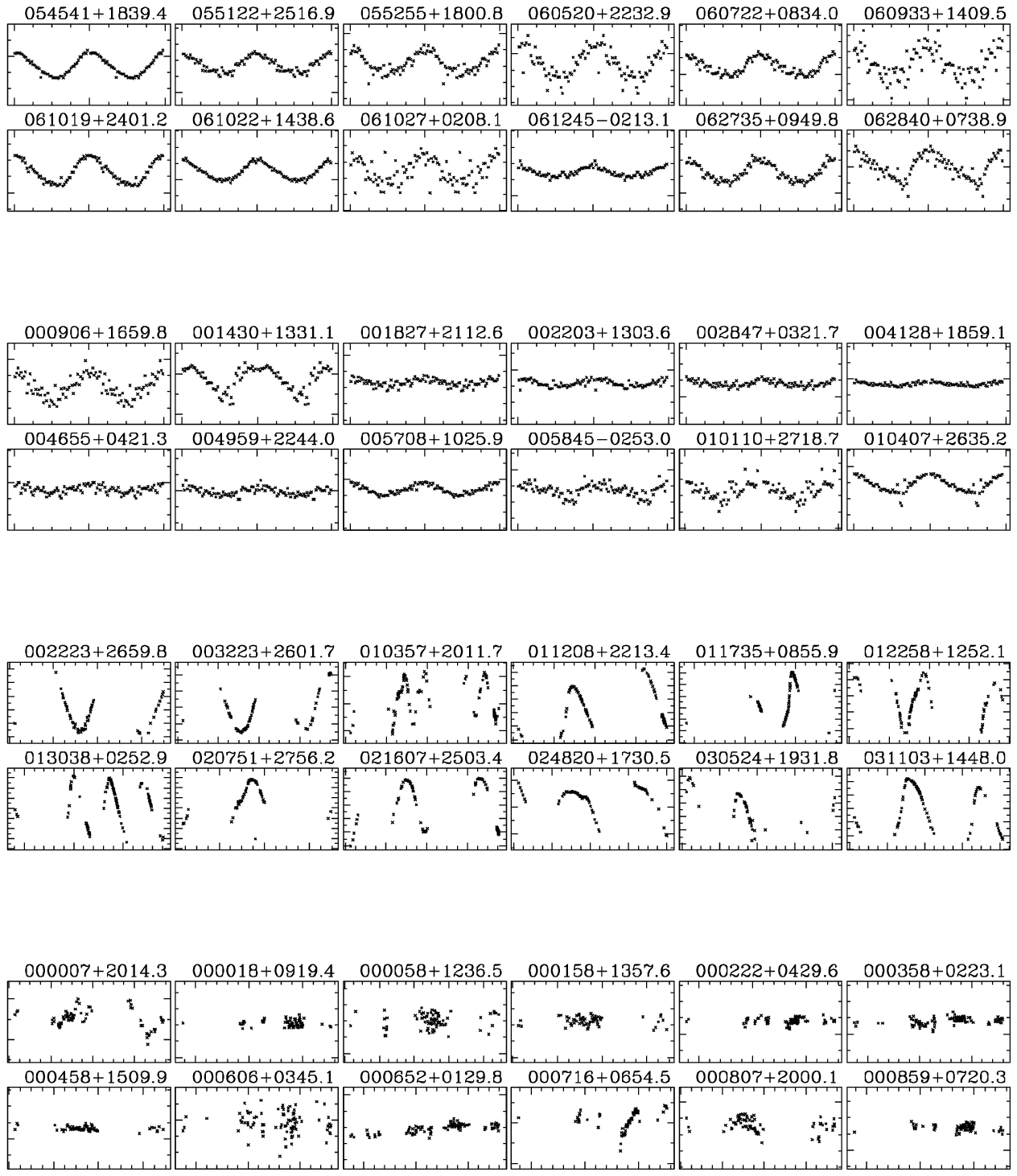}}
\end{figure}

\end{document}